\documentclass[manuscript,aps]{revtex4}
\usepackage{amsmath}
\begin{document}
\title{Energy and momentum of the Friedman and more general universes}
\author{Janusz Garecki}
\affiliation{Institute of Physics, University of Szczecin, Wielkopolska 15; 70-451
Szczecin, POLAND\footnote{e-mail: garecki@sus.univ.szczecin.pl}}
\date{\today}

\begin{abstract}
Recently some authors  concluded that the energy and momentum of the Fiedman universes, flat and closed,
are equal  to zero   locally and globally (flat universes) or only globally (closed universes).
The similar conclusion was also done for more general only homogeneous universes
(Kasner and Bianchi type I).
Such conclusions originated from coordinate dependent calculations performed only in comoving
Cartesian coordinates by using the so-called {\it energy-momentum complexes}.
But it is known that the energy-momentum complexes can be reasonably use only in precisely defined asymptotically
flat spacetimes (at null or at spatial infinity) to calculate global energy and momentum.
In this paper we show, by using new coordinate  independent expressions on energy and momentum
that the Friedman and more general universes {\it needn't be energetic nonentity}.
\end {abstract}
\vspace{3.cm}

\maketitle

\newpage

\section{Introduction}
Recently many authors have calculated the energy and momentum of
the Friedman universes and also more general, only spatially
homogeneous universes, like Kasner, Bianchi type I and Bianchi
type II universes \cite{one}.

The linear elements for these universes read\footnote{We will use
geometrized units in which $G = c=1$.}:
\begin{enumerate}
\item Spatially isotropic and homogeneous Friedman universes in
the coordinates $(t,\chi,\vartheta,\varphi)$
\begin{equation}
ds^2 = dt^2 -R^2(t)[d\chi^2 + S^2(\chi)(d\vartheta^2
+\sin^2\vartheta d\varphi^2)],
\end{equation}
where
\begin{eqnarray}
S(\chi)& =& sin\chi ~~if ~k = 1~(closed ~universe),\nonumber\\
S(\chi)   &=&\chi~~if~ k=0~(flat ~universe),\nonumber\\ ~~
S(\chi) &=& sh\chi ~~if~ k = -1~(open~ universe).
\end{eqnarray}
\item Spatially flat and homogeneous vacuum Kasner's universes in
``Cartesian'' coordinates $(t,x,y,z)$
\begin{equation}
ds^2 = dt^2 - t^{2p_1}dx^2 - t^{2p_2}dy^2 - t^{2p_3}dz^2.
\end{equation}
The constants $p_1,p_2,p_3$ satisfy the following constraints
which follow from the vacuum Einstein equations
\begin{equation}
p_1 + p_2 + p_3 = 1, ~~p_1^2 + p_2^2 + p_3^2 = 1.
\end{equation}
If $p_1 = p_2 = 0, ~p_3 = 1$, then one gets flat Minkowskian
spacetime.
\item Spatially homogeneous Bianchi type I universes filled with stiff matter in
``Cartesian'' coordinates $(t,x,y,z)$
\begin{equation}
ds^2 = dt^2 -e^{2l} dx^2 - e^{2m} dy^2 - e^{2n} dz^2,
\end{equation}
where $l=l(t),~m=m(t),~n = n(t)$.
\item Spatially homogeneous Bianchi type II universes in
``Cartesian'' coordinates $(t,x,y,z)$
\begin{eqnarray}
ds^2 &=& dt^2 -D^2(t)dx^2 - H^2(t)dy^2 -[D^2(t)
+x^2H^2(t)]dz^2\nonumber\\
&-& 2xH^2(t)dydz.
\end{eqnarray}
\end{enumerate}
The above mentioned authors performed their calculations in
special comoving coordinates called ``Cartesian coordinates''
despite that they used {\it coordinate dependent} double index
energy-momentum complexes, matter and gravitation.

The authors have applied the six most frequently used
energy-momentum complexes: Einstein  canonical complex,
Landau-Lifshitz complex, Bergmann-Thomson complex, M\o ller
complex, Papapetrou complex and Weinberg energy-momentum complex.
These all energy-momentum complexes {\it are neither geometrical
objects nor coordinate independent objects}, e.g., they can vanish
in some coordinates locally or globally and in other coordinates
they can be different from zero. It results that the double index
energy-momentum complexes and the gravitational energy-momentum
pseudotensors determined by them {\it have no physical meaning to
a local analysis} of a gravitational field, e.g., to study
gravitational energy distribution. In fact, up to now, they were reasonably
used only to calculate the global quantities for the very
precisely defined asymptotically flat spacetimes (in spatial or in
null direction).

The best one of the all possible double index energy-momentum
complexes from physical and geometrical points of view is the
Einstein canonical double index energy momentum complex $_E K
_i^{~k}$ (For details see, e.g., \cite{Tr62}):
\begin{equation}
_E K_i^{~k}:= \sqrt{\vert g\vert}\bigl(T_i^{~k} + _E
t_i^{~k}\bigr) = {_F U_i^{~[kl]}}_{,l}.
\end{equation}

Here $_F U_i^{~[kl]} = (-) _F U_i^{~[lk]}$ are {\it Freud
superpotentials}, $T^{ik} = T^{ki}$ mean the components of the
symmetric energy-momentum tensor for matter,\footnote{This tensor
is source in the Einstein equations.} and $_E t_i^{~k}$ are the
components of the Einstein canonical {\it energy-momentum
pseudotensor} for gravitational field $\Gamma^i_{~kl} = \bigl\{^i_{kl}\bigr\}$.

From (7) there follow the {\it local conservation laws} for
gravity and matter
\begin{equation}
{_E K_i^{~k}}_{,k}\equiv 0.
\end{equation}
Remark: The other double index energy-momentum complexes have
structure which is very like to the structure of the complex $_E
K_i^{~k}$ and they, of course, also satisfy suitable local conservation laws.

The conclusion of the authors which calculated the energy and
momentum of the Friedman and more general universes by using
double index energy-momentum complexes is the following: the
energy and momentum of the closed Friedman universes {\it are equal to
zero globally}, and in the case of the flat Friedman universes and
their generalizations (Kasner, Bianchi type I, Bianchi type II universes)
these quantities {\it are equal to zero locally and globally}.

One can have at least the following objections against the
calculations of such a kind and against the above conclusion:
\begin{enumerate}
\item The authors despite that they used coordinate dependent
expressions had performed their calculations only in Cartesian
comoving coordinates.

The results obtained in other comoving coordinates, e.g., in
coordinates $(t,\chi,\vartheta,\varphi)$ or in coordinates $(t,r,\vartheta,\varphi)$
{\it are dramatically different}.
\item The local ``energy-momentum distribution'' as given by any
energy-momentum complex {\it has no physical sense} but the
authors try to give a physical sense   of this distribution, e.g.,
they assert that the total energy density for flat Friedman
universes, for Kasner and Bianchi type I universes, is null.
\item The conclusion leads us to Big-Bang which {\it has no
singularity} in total energy density.
\item The global energy and momentum {\it have physical meaning} only
when spacetime is asymptotically flat either in spatial or null
direction and when these quantities can be measured. But this is not a
case of the cosmological models.

So, the problem of the global energy and global linear (or angular) momentum
for Friedman, and for more general universes also, {\it is not
well-posed from the physical point of view} because these
universes are not asymptotically flat spacetimes, and, in
consequence, their global quantities {\it cannot be measurable}.
This problem can only have {\it a mathematical sense}.
\end{enumerate}
 Thus, one can doubt in validity  of the conclusion that the energy
 and momentum of the Friedman, Kasner, Bianchi type I and Bianchi
 type II universes {\it are equal to zero}; especially that all these
 universes {\it are energy-free}.

By using double index energy-momentum complexes one should rather
conclude that the energy and momentum of the Friedman, Kasner,
Bianchi type I, and Bianchi type II universes explicite depend on
the used coordinates and, therefore, they are undetermined {\it not only
locally} but also {\it globally}. The last conclusion is very
sensible because, as we mentioned beforehand, one cannot measure the global
energy and global linear (or angular) momentum of the Friedman and
any more general universe. One can do this only in the case of an
isolated system when spacetime is asymptotically flat.

One cannot use the coordinate independent {\it Pirani} \cite{Tr62} and {\it
Komar} \cite{Gold80}
expressions in order to correctly prove (at least from the
mathematical point of view) the statement that the energy of the
Friedman, Kasner, Bianchi type I and Bianchi type II universes
disappears, i.e., that these universes have zero net energy. It is
because we have no translational timelike Killing vector field
(descriptor of energy in Komar expression) in these universes, and
the privileged normal congruence of the fundamental observers
which exists in these universes is geodesic.\footnote{Pirani
expression on energy only can be applied  in a spacetime having a
privileged normal and timelike congruence. But for a geodesic
congruence Pirani expression fails giving trivially zero.}

One also cannot use for this purpose the coordinate independent
Katz-Bi\v cak-Lynden (BKL)  bimetric approach \cite{BKL96} because
the results obtained in this approach depend on the used
background and on mapping of the spacetime under study onto this
background.

Thus, the ``academic'' statement that the Friedman, Kasner,
Bianchi type I and Bianchi type II universes have no energetic
content {\it is still not satisfactory proved}. But by using Komar
expression, one can correctly (from mathematical point of view)
prove that the linear momentum for these universes disappears in a
comoving coordinates.

In the following we will apply some new expressions on averaged
relative energy and momentum to analyze of the energetic content
of the Friedman, Kasner and Bianchi type I
universes.\footnote{More general universes were not investigated
yet.}

It is interestig that these new expressions lead us to {\it
positive-definite} results for the all these universes.

\section{Using of the averaged relative energy-momentum tensors to
analyze energy and momentum of the Friedman and more general
homogeneous universes}

In the papers \cite{Gar8} we have defined the canonical superenergy and
angular supermomentum tensors, matter and gravitation, in general
relativity ({\bf GR}) and studied their properties and physical
applications. In the case of the gravitational field these tensors gave
us some substitutes of the non-existing gravitational
energy-momentum and gravitational angular momentum tensors.

The canonical superenergy and angular supermomentum tensors were
obtained pointwise as a result of some special averaging of the
differences of the energy-momentum and angular momentum in normal
coordinates {\bf NC(P)}. The role of the normal coordinates {\bf NC(P)}
is, of course, auxilliary, only to extract tensorial quantities even
from pseudotensorial ones.

The dimensions of the components of the canonical superenergy and
angular supermomentum tensors can be written
down as:[the dimensions of the components of an energy-momentum or
angular momentum tensor (or pseudotensor)]$\times m^{-2}$.

Recently, in the paper \cite{Gar9} we have proposed a new averaging of the energy-momentum and
angular momentum differences in {\bf NC(P)} which is very like to the
averaging used in \cite{Gar8} and which gives the averaged
quantities with proper dimensionality of the energy-momentum and angular
momentum densities.

Namely, we have proposed the following general definition of the averaged tensor
(or pseudotensor)  $T_a^b$
\begin{equation}
<{\it T_a^{~b}}(P)> := \displaystyle\lim_{\varepsilon\to
0}{\int\limits_{\Omega}{\bigl[{\it T_{(a)}^{~~~(b)}}(y) - {\it
T_{(a)}^{~~~(b)}}(P)\bigr]
d\Omega}\over\varepsilon^2/2\int\limits_{\Omega}d\Omega},
\end{equation}
where
\begin{equation}
{\it T_{(a)}^{~~~(b)}}(y) := T_i^{~k}(y){}e^i_{~(a)}(y){}e_k^{~(b)}(y),
\end{equation}
\begin{equation}
{\it T_{(a)}^{~~~(b)}}(P):= T_i^{~k}(P){}e^i_{~(a)}(P){}e_k^{~(b)}(P) = {\it T_a^{~b}}(P)
\end{equation}
are the tetrad (or physical) components of a tensor or a pseudotensor
$T_i^{~k}(y)$ which describes an energy-momentum distribution, $y$ is the collection of normal coordinates {\bf NC(P)}
at a given point {\bf P}, $e^i_{~(a)}(y),~e_k^{~(b)}(y)$ denote an
orthonormal tetrad field and its dual, respectively,
\begin{equation}
e^i_{~(a)}(P) = \delta^i_a,~e_k^{~(a)}(P)
=\delta^a_k,~e^i_{~(a)}(y)e_i^{~(b)}(y) =\delta_a^b,
\end{equation}
and they are parallelly propagated along geodesics through {\bf P}.

For a sufficiently small domain $\Omega$ which surrounds {\bf P} we
required
\begin{equation}
\int\limits_{\Omega}{y^id\Omega} = 0,~~\int\limits_{\Omega}{y^iy^kd\Omega} =
\delta^{ik} M,
\end{equation}
where
\begin{equation}
M = \int\limits_{\Omega}{(y^0)^2 d\Omega} = \int\limits_{\Omega} {(y^1)^2
d\Omega} = \int\limits_{\Omega}{(y^2)^2 d\Omega} =
\int\limits_{\Omega}{(y^3)^2 d\Omega},
\end{equation}
is a common value of the moments of inertia of the domain $\Omega$ with
respect to the subspaces $y^i = 0, ~~(i = 0,1,2,3)$.

We have choosen $\Omega$ as a small analytic ball defined by
\begin{equation}
(y^0)^2 + (y^1)^2 + (y^2)^2 + (y^3)^2 \leq R^2 = \varepsilon^2 L^2,
\end{equation}
which can be described in a covariant way in terms of the auxiliary
positive-definite metric $h^{ik} := 2v^iv^k - g^{ik}$, where $v^i$ are
the components of the four-velocity of an observer {\bf O} at rest at
{\bf P} (See, e.g., \cite{Gar8} ). $\varepsilon >0$ means an undimensional
parameter, and $L >0$ is a fundamental length.

The mathematical trick with putting  $R=\varepsilon L$ was
discovered by B. Mashhoon  \cite{Mas10} . This trick leads to proper
dimensionality of the averaged quantities.

Since at {\bf P} the tetrad and normal components are equal, from now on
we will write the components of any quantity at {\bf P} without (tetrad)
brackets, e.g., $T_a^{~b}(P)$ instead of $T_{(a)}^{~~~(b)}(P)$ and so
on.

For the matter energy-momentum tensor $_m T_a^{~b}(y)$ the
averaging formula (9)  gives
\begin{equation}
<_m T_a^{~b}(P)> = _m S_a^{~b}(P) {L^2\over 6},
\end{equation}
where
\begin{equation}
_m S_a^{~b}(P) := \delta^{lm}\nabla_{(l}\nabla_{m)}{\hat T}_a^{~b}
\end{equation}
is the {\it canonical superenergy tensor of matter} \cite{Gar8}.

By introducing the four velocity ${\hat v}^l \dot = \delta^l_0,~v^lv_l =1$ of
an observer {\bf O} at rest at {\bf P} and the local metric ${\hat
g}^{ab}\dot = \eta^{ab}$, where $\eta^{ab}$ is the inverse Minkowski
metric, one can write (17) in a covariant way as
\begin{equation}
_m S_a^{~b} (P;v^l) = \bigl(2{\hat v}^l{\hat v}^m -
{\hat g}^{lm}\bigr)\nabla_{(l}\nabla_{m)} {\hat T}_a^{~b}.
\end{equation}
The sign $\dot =$ means that an equality is valid only in some special
coordinates.

For the gravitational field one gets the following tensor (if one uses the Einstein
canonical energy-momentum pseudotensor $_E t_a^{~b}(y)$ in the
averaging process)
\begin{equation}
<_g t_a^{~b}(P;v^l)> = _g S_a^{~b}(P;v^l){L^2\over 6},
\end{equation}
where the tensor $_g S_a^{~b}(P;v^l)$ is the {\it canonical superenergy
tensor} for the gravitational field \cite{Gar8}.

We have \cite{Gar8}
\begin{eqnarray}
_g S_a^{~b}(P;v^l) &=& {2\alpha\over 9}\bigl(2{\hat v}^l{\hat v}^m -
{\hat g}^{lm}\bigr)\biggl[{\hat B}^b_{~alm} + {\hat
P}^b_{~alm}\nonumber\\
&-& 1/2\delta_a^b{\hat R}^{ijk}_{~~~m}\bigl({\hat R}_{ijkl} + {\hat
R}_{ikjl}\bigr) + 2\beta^2\delta_a^b{\hat E}_{(l\vert g}{\hat
E}^g_{~\vert m)}\nonumber\\
&-& 3\beta^2{\hat E}_{a(l\vert}{\hat E}^b_{~\vert m)} + 2\beta{\hat
R}^b_{~(ag)(l\vert} {\hat E}^g_{~\vert m)}\biggr].
\end{eqnarray}

Here $\alpha = {1\over 16\pi} = {1\over 2\beta}$  and $E_i^{~k} := T_i^{~k} - 1/2 T_a^a$
is the modified energy-momentum tensor of matter. $B^b_{~alm}$
mean the components of the Bel-Robinson tensor and $P^b_{~alm}$
are the components of a tensor which is very closely related to
the Bel-Robinson tensor.\footnote{$P^b_{~alm}$ has almost the same
analytic form as $B^b_{~alm}$ and the same symmetries.}

In vacuum the tensor $<_g t_a^{~b}(P;v^l)>$ reduces to the simpler form
\begin{equation}
<_g t_a^{~b} (P;v^l)> = {4\alpha\over 27}\bigl(2{\hat v}^l{\hat v}^m -
{\hat g}^{lm}\bigr)\biggl[{\hat R}^{b(ik)}_{~~~~~(l\vert}{\hat
R}_{aik\vert m)} -1/2\delta_a^b{\hat R}^{i(kp)}_{~~~~~(l\vert}{\hat
R}_{ikp\vert m)}\biggr]L^2,
\end{equation}
which is symmetric and the quadratic form $<_g t_{ab}(P;v^l)>{\hat
v}^a{\hat v}^b$ is {\it positive-definite}.

The averaged energy-momentum tensors $<_m T_a^{~b}(P;v^l)>$ and $<_g
t_a^{~b}(P;v^l)>$ can be considered as the {\it averaged tensors of the relative
energy-momentum}.

The averaged tensors
\begin{equation}
<_g t_a^{~b}(P;v^l)>, ~<_m T_a^{~b}(P;v^l)>,
\end{equation}
depend on the four-velocity ${\vec v}$ of a fiducial observer {\bf O} which is at rest at
the beginning {\bf P} of the normal coordinates {\bf NC(P)} used for
averaging and on some fundamental length $L>0$.

One can try to fix the fundamental length $L$ in a some way, e.g., with the help  of the loop quantum gravity
({\bf LQG}). Namely, one can take as $L$ the smallest length over which the classical model
of the spacetime is admissible. But this is not necessary. One can effectively use the averaged
relative energy-momentum tensors without fixing $L$ explicitly (See  \cite{Gar9} for details).

After fixing the fundamental length $L$ one can determine univocally the averaged relative
energy-momentum tensors tensors along the world line of an observer {\bf O}. In general one can
{\it unambiguously determine} these tensors (after fixing $L$) in the whole spacetime or in some
domain $\Omega$ if in the spacetime or in the domain $\Omega$ a geometrically distinguised
timelike unit vector field ${\vec v}$ exists. An example of such a kind of the spacetime is
given by universes considered in this paper.

Let us apply the averaged relative energy-momentum tensors for gravitation\hfill\break
$<_g t_i^{~k}(P;v^l)>$ and for matter $<_m T_i^{~k}(P;v^l)>$ to analyze the Friedman and more
general universes (Kasner vacuum universes and Bianchi type I universes filled with stiff matter).

With this aim let us define
\begin{equation}
_g\epsilon := <_g t_a^{~b}(p;v^l)>v_av^b
\end{equation}
----- the averaged relative gravitational energy density,
\begin{equation}
_m\epsilon:= <_m T_a^{~b}(P;v^l)>v^av_b
\end{equation}
----- the averaged relative matter energy density,
and
\begin{equation}
\epsilon:= _g\epsilon + _m \epsilon
\end{equation}
----- the averaged relative total energy density.

Here $v^a$ are the components of the four-velocity of an observer {\bf
O} which is studying gravitational and matter fields.

If we take as the observers {\bf O} the globally
defined set of the fundamental observers,\footnote{For these observers $v^a = \delta^a_o$
in a comoving coordinates.} then we can also define the
global averaged total relative energy $E$ for the
considered universes:

\begin{equation}
E := \int\limits_{t = const}\epsilon\sqrt{\vert g\vert} d^3v,
\end{equation}
and, in analogous way, the global averaged relative energy for matter
and for gravitation.

Here $d^3v$ means the product of the differentials of
the coordinates which parametrize slices $t= const$, e.g., $d^3v = dxdydz$
in the Cartesian comoving coordinates $(t,x,y,z)$.

After something tedious but very simple calculations we will obtain for Friedman universes:
\begin{enumerate}
\item $_g\epsilon$, $_m\epsilon$ and, in consequence $\epsilon$, are
{\it positive definite} for the all Friedman universes.
\item $\displaystyle\lim_{R\to 0}{}_g\epsilon = \displaystyle\lim_{R\to 0}
{}_m\epsilon =\displaystyle\lim_{R\to 0}{}\epsilon = +\infty, ~~(k =
0,^+_- 1)$.

It follows from this that one can use the averaged relative energy densities to study the
Big-Bang singularity.
\item $\displaystyle\lim_{R\to\infty}{} _g\epsilon =
\displaystyle\lim_{R\to\infty}{} _m \epsilon =
\displaystyle\lim_{R\to\infty}{}\epsilon  = 0, ~~(k = 0,-1)$.
\item The global averaged relative energies, gravitation, matter and total, are infinite
($+\infty$) for flat and for open Friedman universes and they are finite and positive
for closed Friedman universes.
\end{enumerate}

For vacuum Kasner universes and for expanding Bianchi type I
universes filled with stiff matter one obtains the following,
coordinate independent results:
\begin{enumerate}
\item The averaged relative gravitational energy of a vacuum
Kasner universe {\it has positive-definite density} and the same
limits when $t\longrightarrow 0^+$ or when $t\longrightarrow \infty$
as in the case of a flat Friedman universe. Also the global
averaged relative energy is divergent to $+\infty$.
\item For an expanding Bianchi type I universe filled with stiff matter
the averaged relative energy densities, gravitation and matter,
are still positive definite and lead to divergent to $+\infty$
global energies.
\end{enumerate}

The other three invariant integrals which formally represent the components
$P_{(\alpha)} ~~(\alpha = 1,2,3)$ of the global averaged relative linear momentum for
Friedman and for more general, only homogeneous, universes
\begin{equation}
P_{(\alpha)}:= \int\limits_{t = const}\bigl\{<_g t_i ^{~0}> + < _m
T_i ^{~0}>\bigr\}e^i_{~(\alpha)}\sqrt{\vert g\vert}d^3v,~~(\alpha = 1,2,3),
\end{equation}
{\it vanish trivially} in a comoving coordinates  because the
integrands in these integrals (densities of the averaged relative linear
momentum components) {\it identically vanish}.

Here $e^i_{~(\alpha)}, ~~(\alpha = 1,2,3)$ mean the components of the
three translational spatial Killing vector fields (descriptors of the
linear momentum) which exist in the Friedman universes and in the more general, only homogeneous,
universes (See, e.g., \cite{Gar10}).

We would like to emphasize that the integrals (26) and (27) do not
depend on the used coordinates. They depend only on a slice $t = const$.

The all above results are very sensible and satisfactory from the
physical point of view.

\section{Conclusion}

The new, coordinate independent expressions on the averaged relative energy-momentum
indicate that the Friedman, Kasner and Bianchi type I universes
{\it are not energetic nonentity}: all they have {\it positive-definite} averaged
relative energy densities.

The result of such a kind is very satisfactory from the physical
point of view. Much more satisfactory than the strange and
coordinate dependent results which one obtains when uses the double
index energy-momentum complexes and pseudotensors.

We are planning to use in a near future the averaged relative
energy-momentum tensors, and also the averaged tensors of the relative
angular momentum,\footnote{One can introduce such tensors in analogy
to the averaged relative energy-momentum tensors. See  \cite{Gar9}   for details.}
to analyze more general homogeneous universes than universes considered in
this  paper.

\appendix
\section{The averaged relative energy densities for gravity and
for matter in Kasner and in Bianchi type I universes}

Here we give the final expression for $_g\epsilon$
in Kasner vacuum universes and $_g\epsilon$ and $_m\epsilon$ in Bianchi type I universes filled
with stiff matter. The corresponding expressions for Friedman
universes can be easily found from the results presented in the old our papers \cite{Gar8}
in which we have studied superenergy of the Friedman universes
and from the formulas  (16) and  (19) of this paper which connect
the averaged relative energy-momentum tensors with the canonical superenergy tensors.
\begin{enumerate}
\item $_g\epsilon$ for gravity in a Kasner universe:
\begin{equation}
_g\epsilon = {2\alpha\over 9t^4}\bigl(p_1^2 p_2^2 + p_1^2 p_3^2
+p_2^2 p_3^2\bigr)L^2 + {2\alpha\over 27t^4}\bigl[p_1^2(p_1 - 1)^2 +
p_2^2(p_2 -1)^2 + p_3^2(p_3 - 1)^2\bigr]L^2.
\end{equation}
\item $_g\epsilon$ and $_m\epsilon$ in an expanding Bianchi type I
universe filled with stiff matter:
\begin{eqnarray}
_g\epsilon &=& {2\alpha\over 27}\bigl\{ [({\dot l})^2 + {\ddot l}]^2
+[({\dot m})^2 + {\ddot m}]^2 +[({\dot n})^2 + {\ddot
n}]^2\nonumber\\
&+& 2\bigl({\dot l}^4 {\dot m}^4 + {\dot l}^4 {\dot n}^4 + {\dot
m}^4 {\dot n}^4\bigr) + {\dot m}^2 {\dot l}^2 + {\dot m}^2 {\dot
n}^2 + {\dot l}^2 {\dot n}^2\bigr\}L^2\nonumber\\
&+& {20\alpha\over 27}\bigl({\dot l}{\dot m} + {\dot l}{\dot n} +
{\dot m}{\dot n}\bigr)^2 L^2,
\end{eqnarray}
\begin{eqnarray}
_m\epsilon &=& {\alpha\over 3}\bigl[{\dddot l}\bigl({\dot m} + {\dot
n}\bigr) + {\dddot m}\bigl({\dot l}+ {\dot n}\bigr) + {\dddot
n}\bigl({\dot l} + {\dot m}\bigr)  + 2\bigl({\ddot l}{\ddot m} +
{\ddot l}{\ddot n} + {\ddot m} {\ddot n}\bigr)\bigr]L^2\nonumber\\
&+&{4\alpha\over 3}\bigl({\dot l}{\dot m} + {\dot l}{\dot n} +
{\dot m}{\dot n}\bigr)\bigl({\dot l}^2 + {\dot m}^2 + {\dot
n}^2\bigr)L^2\nonumber\\
&-& {\alpha\over 3}\bigl[{\ddot l}\bigl({\dot m} + {\dot n}\bigr) + {\ddot
m}\bigl({\dot l} + {\dot n}\bigr) + {\ddot n}\bigl({\dot l} + {\dot m}\bigr)\bigr]
\bigl({\dot l} + {\dot m} + {\dot n}\bigr)L^2.
\end{eqnarray}
\end{enumerate}
It is immediately seen that the averaged relative gravitational energy
densities are in the both above cases {\it positive definite}.

With the help of the Einstein equations one can easily prove that
the averaged relative energy density for matter in an expanding  Bianchi
type I universe {\it is also positive-definite}.

\end{document}